\documentclass{emulateapj}
\usepackage{natbib}

\newcommand{\myemail}{lacerda.pedro@gmail.com}
\newcommand{\tna} {\,\tablenotemark{a}}
\newcommand{\tnb} {\,\tablenotemark{b}}

\newcommand{\mum} {$\mu$m}

\slugcomment{Accepted for publication, The Astronomical Journal.  2008 November 28.}
\shorttitle{Near-IR Photometry of Haumea}
\shortauthors{Pedro Lacerda}

\def\FigLightcurve{
   \begin{figure}
   \centering
      \includegraphics[width=0.5\textwidth]{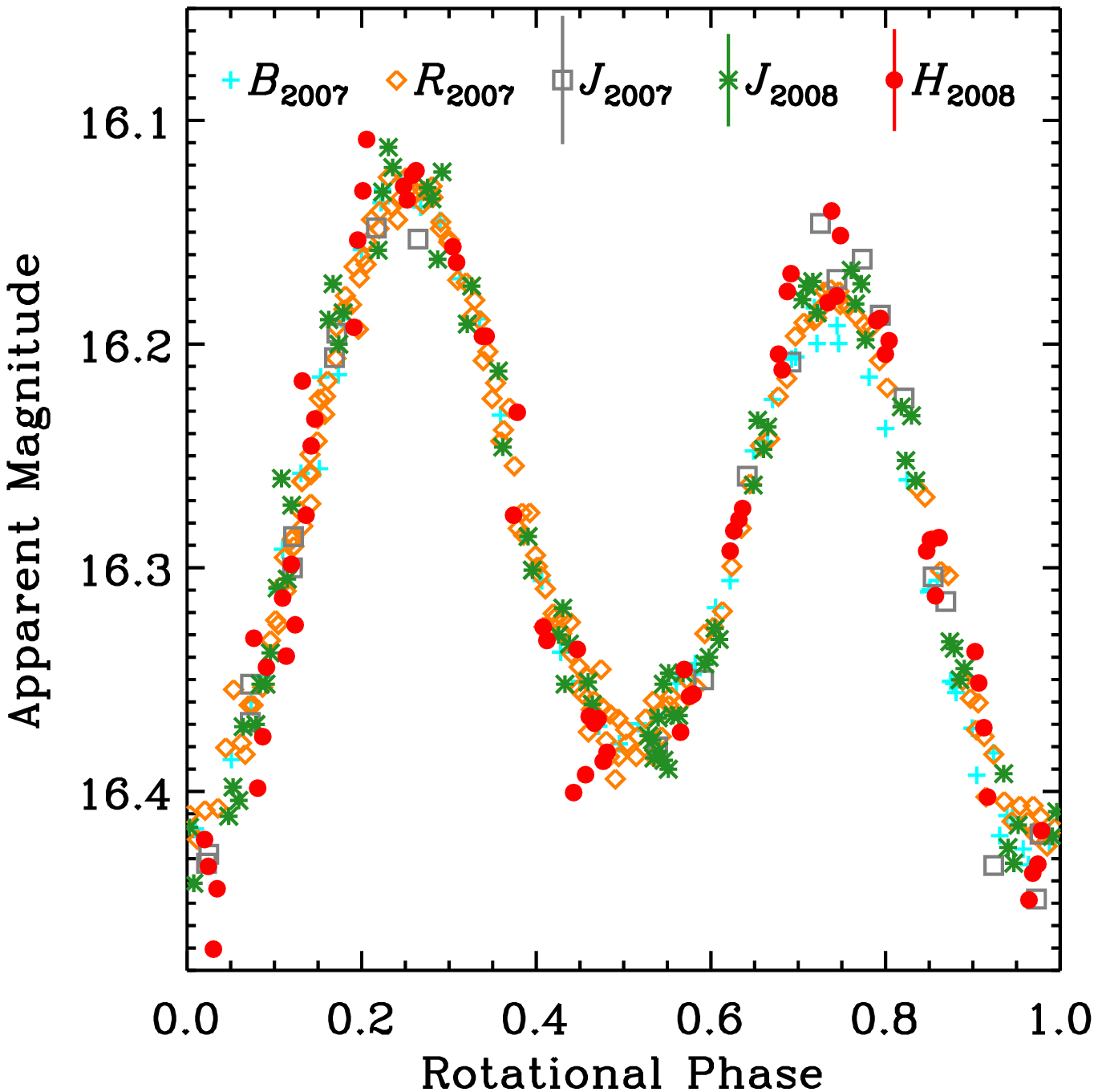}

   \caption[lightcurves] {Four-color lightcurve of Haumea.  Symbols labelled
2008 correspond to the new $J$ and $H$ data presented here.  Measurements from
2007 \citep{2008AJ....135.1749Lac} are marked for direct comparison.  The
vertical axis corresponds to the $J_{2008}$ apparent magnitude.  Data in other
bands have been shifted using the mean colors (see text).} 

   \label{Fig.Lightcurve}
   \end{figure}
}

\def\FigLightcurveBinned{
   \begin{figure}
   \centering
      \includegraphics[width=0.5\textwidth]{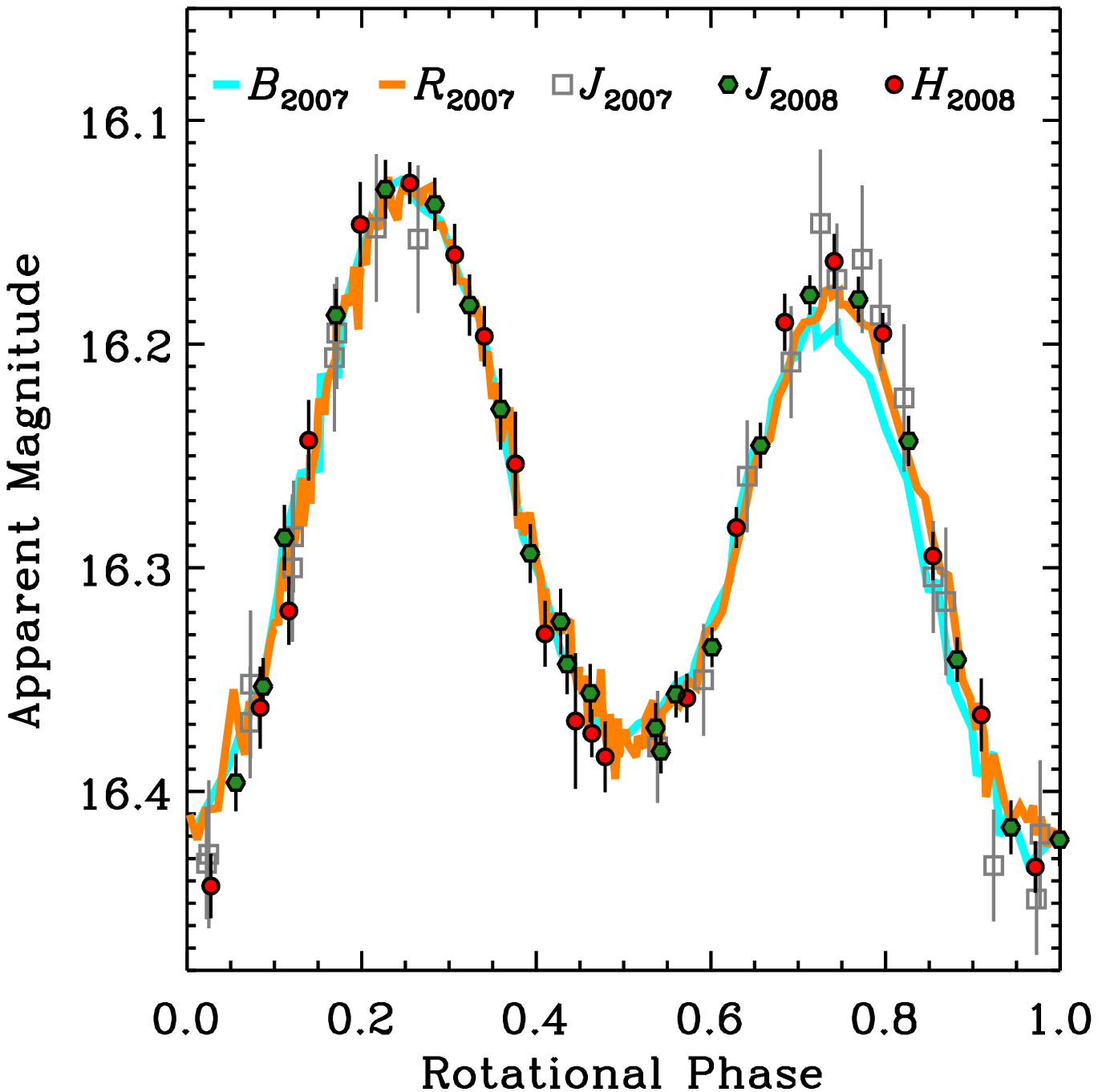}

   \caption {Binned near-infrared lightcurve of Haumea.  Here, sets of
consecutive data points in Fig.~\ref{Fig.Lightcurve} have been binned to reduce
scatter.  Green hexagons ($J$-band) and red circles ($H$-band) mark the mean
rotational phase and magnitude of consecutive measurements (mostly sets of
four).  Measurements from 2007 \citep{2008AJ....135.1749Lac} are marked for
direct comparison: the green squares are $J$-band measurements, while the
densely sampled $B$- and $R$-band data are plotted as thick cyan and orange
lines.  The vertical axis corresponds to the $J_{2008}$ apparent magnitude.
Data in other bands have been shifted using the mean colors (see text).} 

   \label{Fig.LightcurveBinned}
   \end{figure}
}

\def\FigColorcurves{
   \begin{figure}
   \centering
      \includegraphics[width=0.46\textwidth]{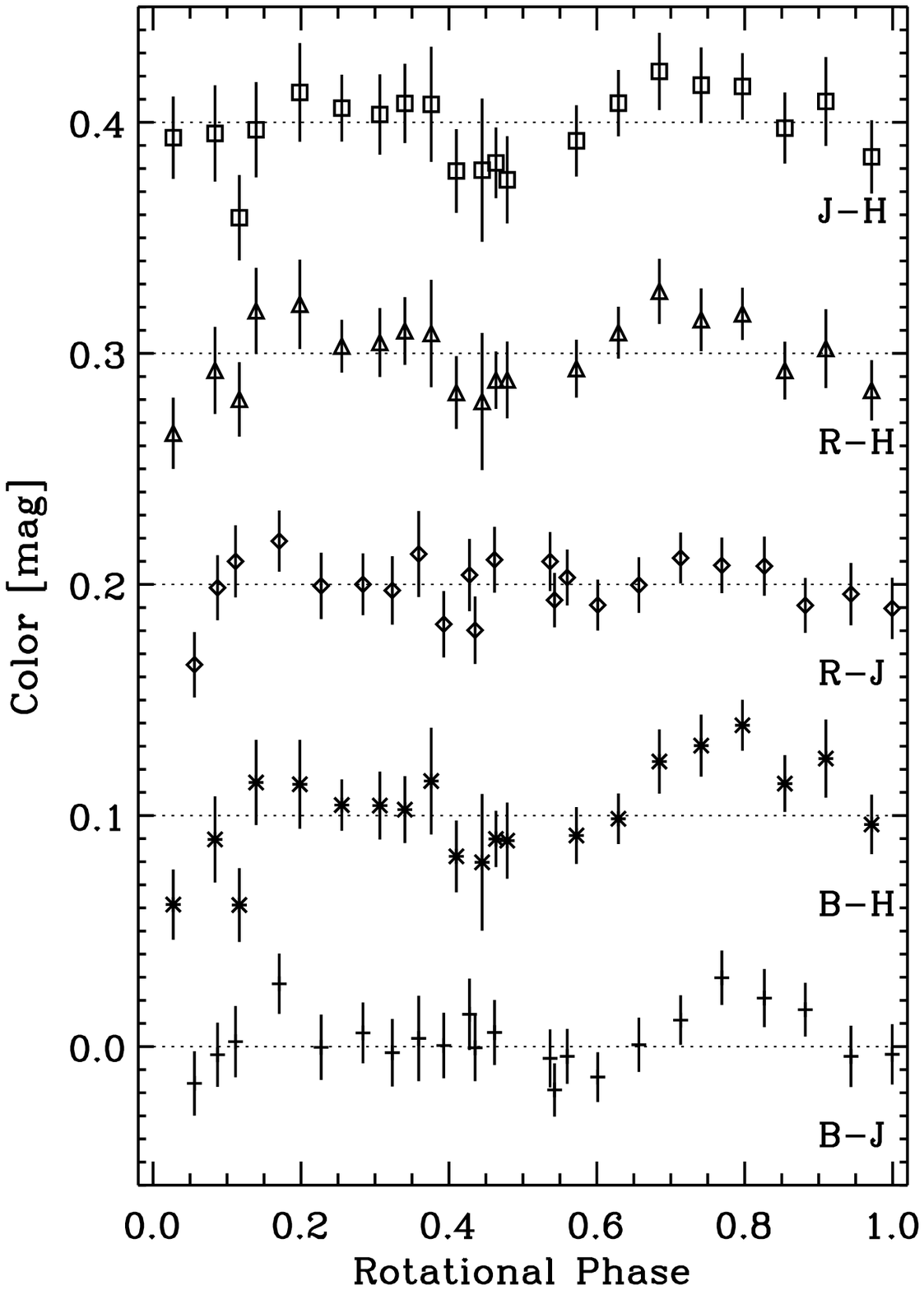}

   \caption {Visible and near-infrared color curves of Haumea.  Each curve
represents deviations from the respective rotationally-medianed mean color
(from bottom up, $B$$-$$J=1.856\pm0.012$ mag,  $B$$-$$H=1.799\pm0.021$ mag,
$R$$-$$J=0.885\pm0.012$ mag, $R$$-$$H=0.828\pm0.016$ mag, and
$J$$-$$H=-0.057\pm0.016$ mag) and is vertically shifted by 0.1 mag for clarity.
Slight reddening bumps are observable in $B$$-$$H$, $B$$-$$J$, and possibly in
$R$$-$$H$ and $J$$-$$H$.} 

   \label{Fig.ColorCurves}
   \end{figure}
}

\def\FigGaussianProb{
   \begin{figure}
   \centering
      \includegraphics[width=0.46\textwidth]{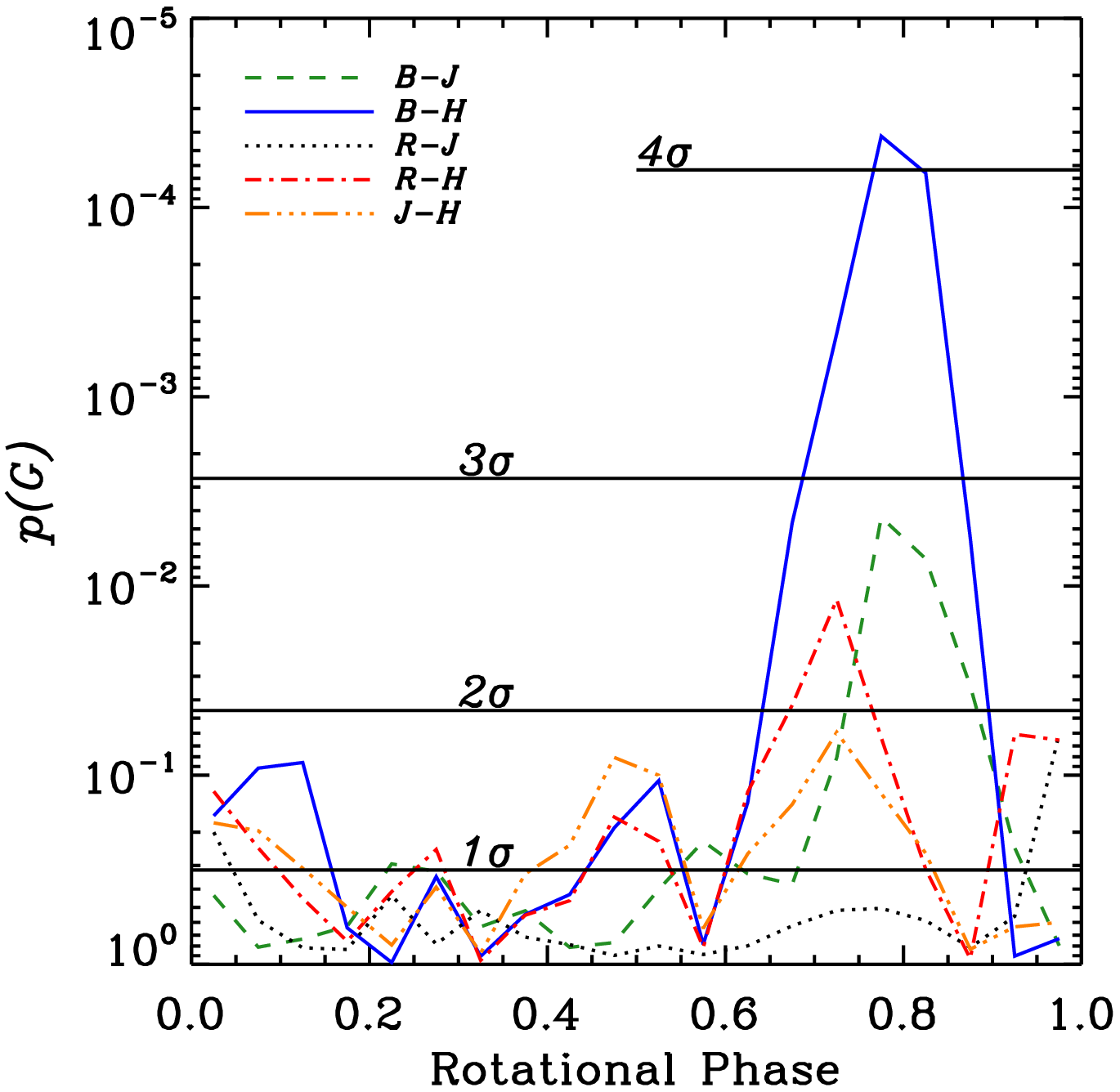}

   \caption {Running Gaussian probability for each of the color curves in
Fig.~\ref{Fig.ColorCurves}.  Unlikely sequences of points all above or below
the median colors (horizontal, dotted lines in Fig.~\ref{Fig.ColorCurves}) will
be visible as peaks in this Figure.  The color curve $B$$-$$H$ color shows a
significant peak at rotational phase $\phi\sim0.8$. Curves $B$$-$$J$ and
$R$$-$$H$ show smaller but noticeable features at the same rotational phase.} 

   \label{Fig.GaussianProb}
   \end{figure}
}

\def\FigGaussianProbVsWindow{
   \begin{figure}
   \centering
      \includegraphics[width=0.46\textwidth]{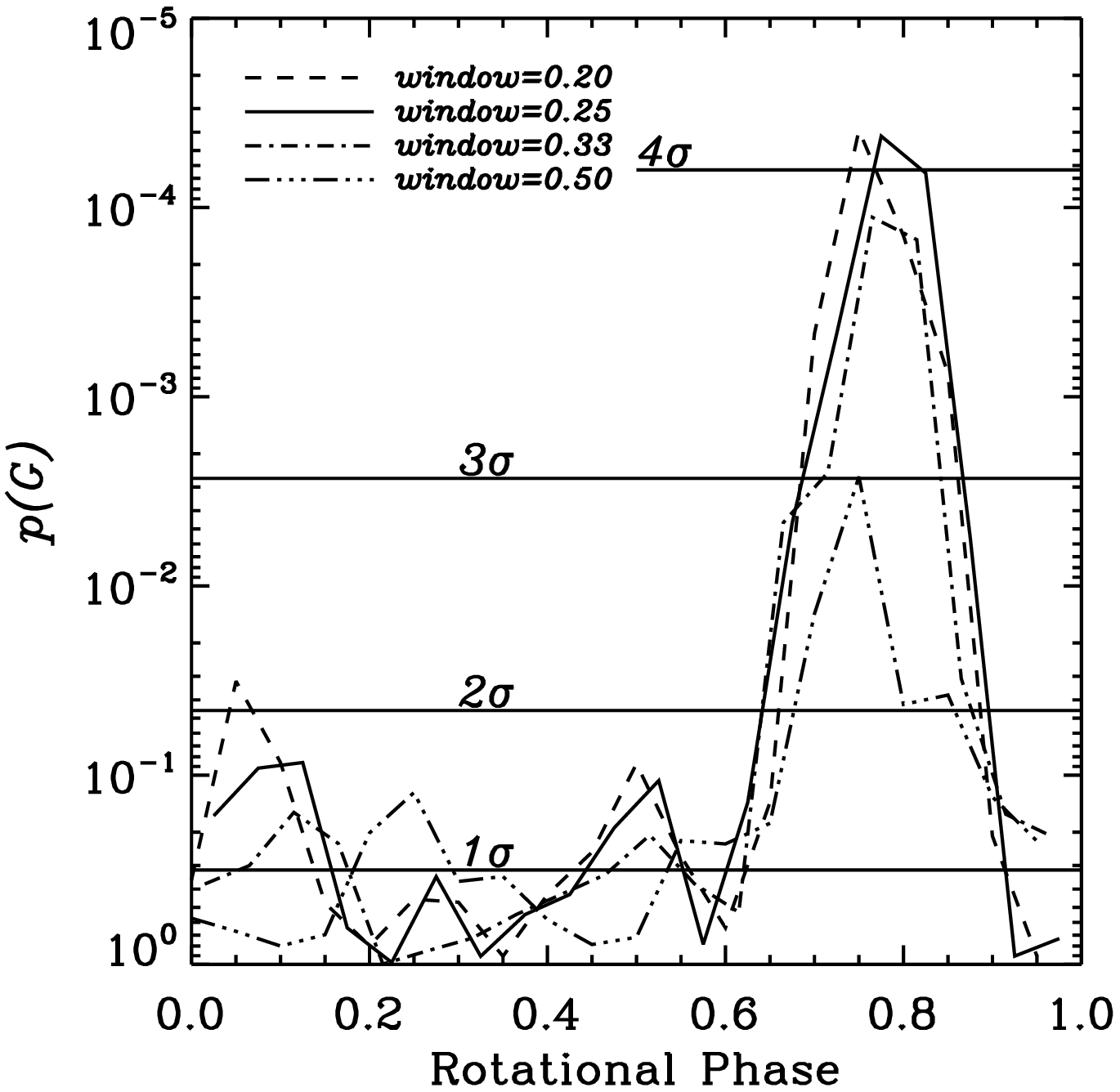}

   \caption {Running Gaussian probability for the $B$$-$$H$ color curve for
four window sizes, $\Delta\phi=0.20$, 0.25, 0.33, and 0.50.  The running
probability $p(G)$ is calculated in a window of width $\Delta\phi$ which is
evaluated in rotational phase steps of 0.05.  The differences are minimal for
window sizes $0.20\leq\Delta\phi\leq0.33$.  For the largest window size
$\Delta\phi=0.50$ the probability begins to appear dilluted.  See text for
details.} 

   \label{Fig.GaussianProbVsWindow}
   \end{figure}
}

\def\FigSpectrumVsPhase{
   \begin{figure}
   \centering
      \includegraphics[width=0.46\textwidth]{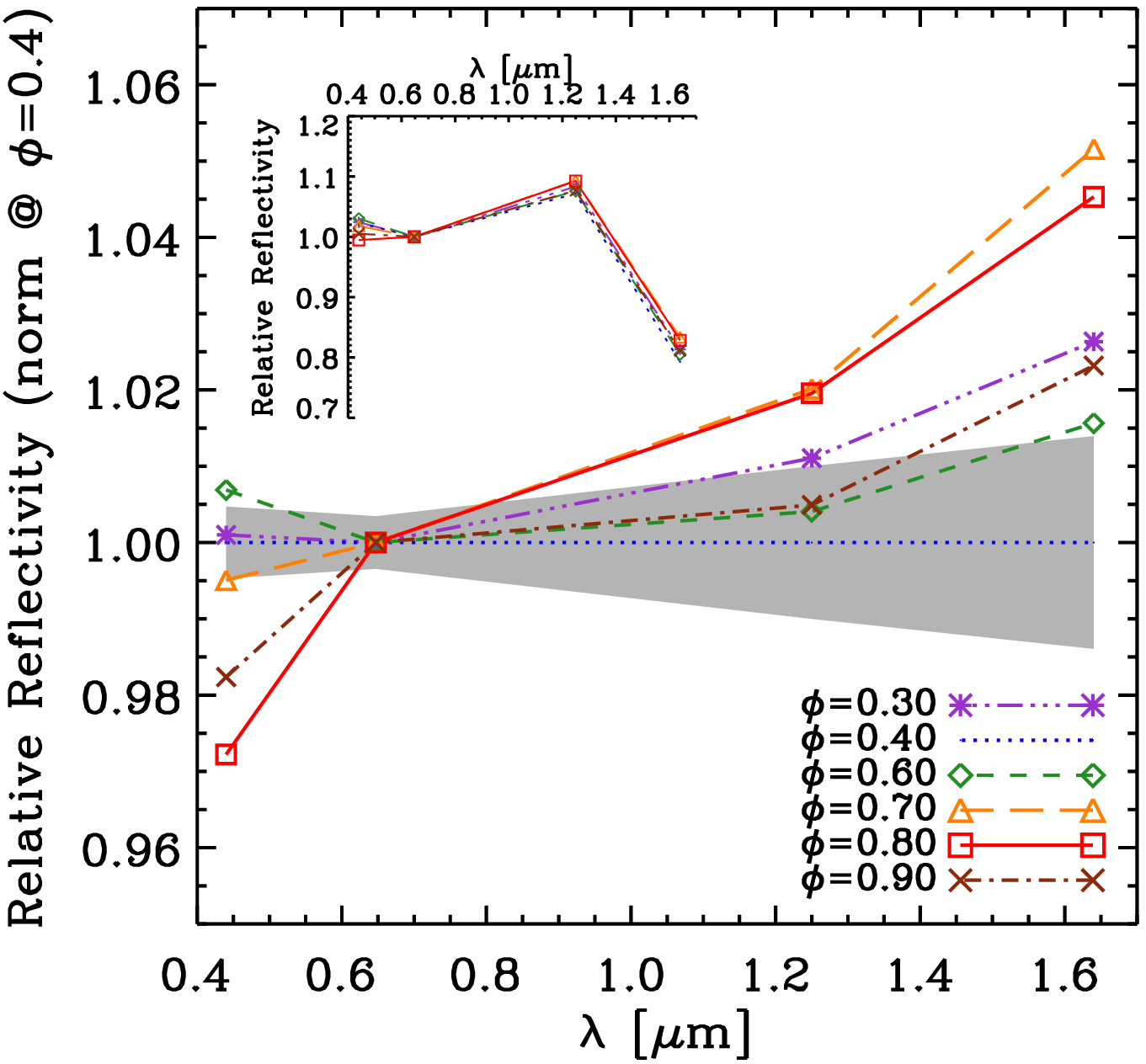}

   \caption {Normalized reflectivity vs.\ wavelength for different rotational
phases.  The reflectivities were computed from the broadband colors by
subtracting the colors of the Sun \citep[$B-R=0.996$, $R-J=0.797$, $J-H=0.258$;
cf.][]{2006MNRAS.367..449Hol}.  All curves were normalized at the $R$-band, and
divided by the curve at rotational phase $\phi=0.4$, for clarity.  The inset
shows the reflectivities prior to dividing by the $\phi=0.4$ curve.  The shaded
polygon represents the uncertainty propagated from errors of measurement.} 

   \label{Fig.SpectrumVsPhase}
   \end{figure}
}

\def\FigPhaseCoefficients{
   \begin{figure}
   \centering
      \includegraphics[width=0.46\textwidth]{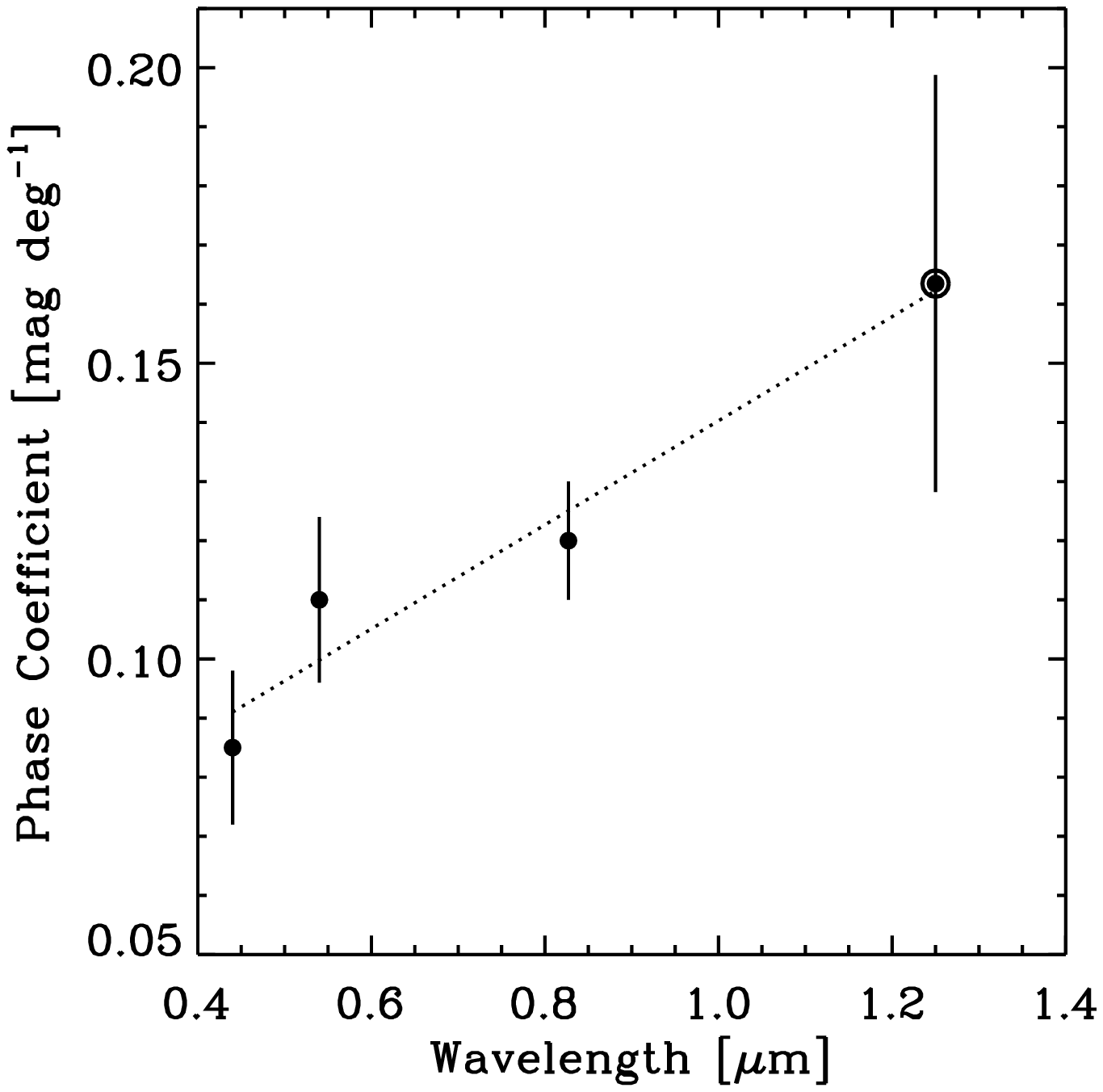}

   \caption {Phase coefficient of Haumea as a function of wavelength.  From
left to right, points mark the central wavelength of the $B$, $V$, $I$, and $J$
bands.  Our measurement of the $J$-band phase coefficient,
$\beta_J=0.16\pm0.04\,$mag$\,$deg$^{-1}$ is highlighted.  The remaining three
points are taken from \citet{2006ApJ...639.1238R}.  A linear best-fit (slope
0.09 and 1 \mum\ value 0.14) is overplotted as a dotted line.} 

   \label{Fig.PhaseCoefficients}
   \end{figure}
}

\def\FigDBetaDLambdaVsAlbedo{
   \begin{figure}
   \centering
      \includegraphics[width=0.46\textwidth]{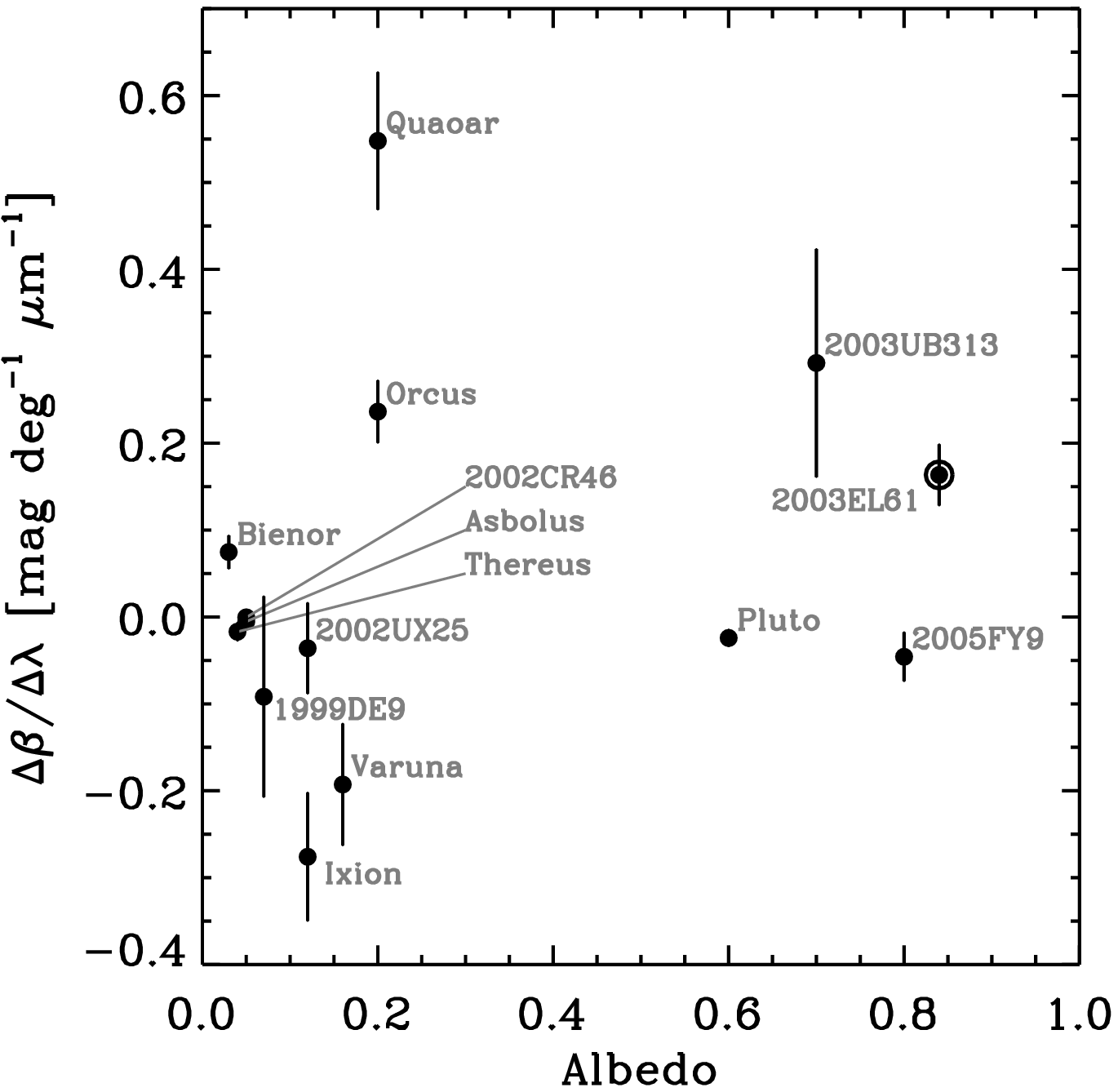}

   \caption {Slope of the $\beta(\lambda)$ relation versus approximate
geometric albedo for a number of KBOs and Centaurs.  No clear relation exists
between the two quantities.} 

   \label{Fig.DBetaDLambdaVsAlbedo}
   \end{figure}
}

\def\FigSatelliteJ{
   \begin{figure*}
   \centering
      \includegraphics[width=0.8\textwidth]{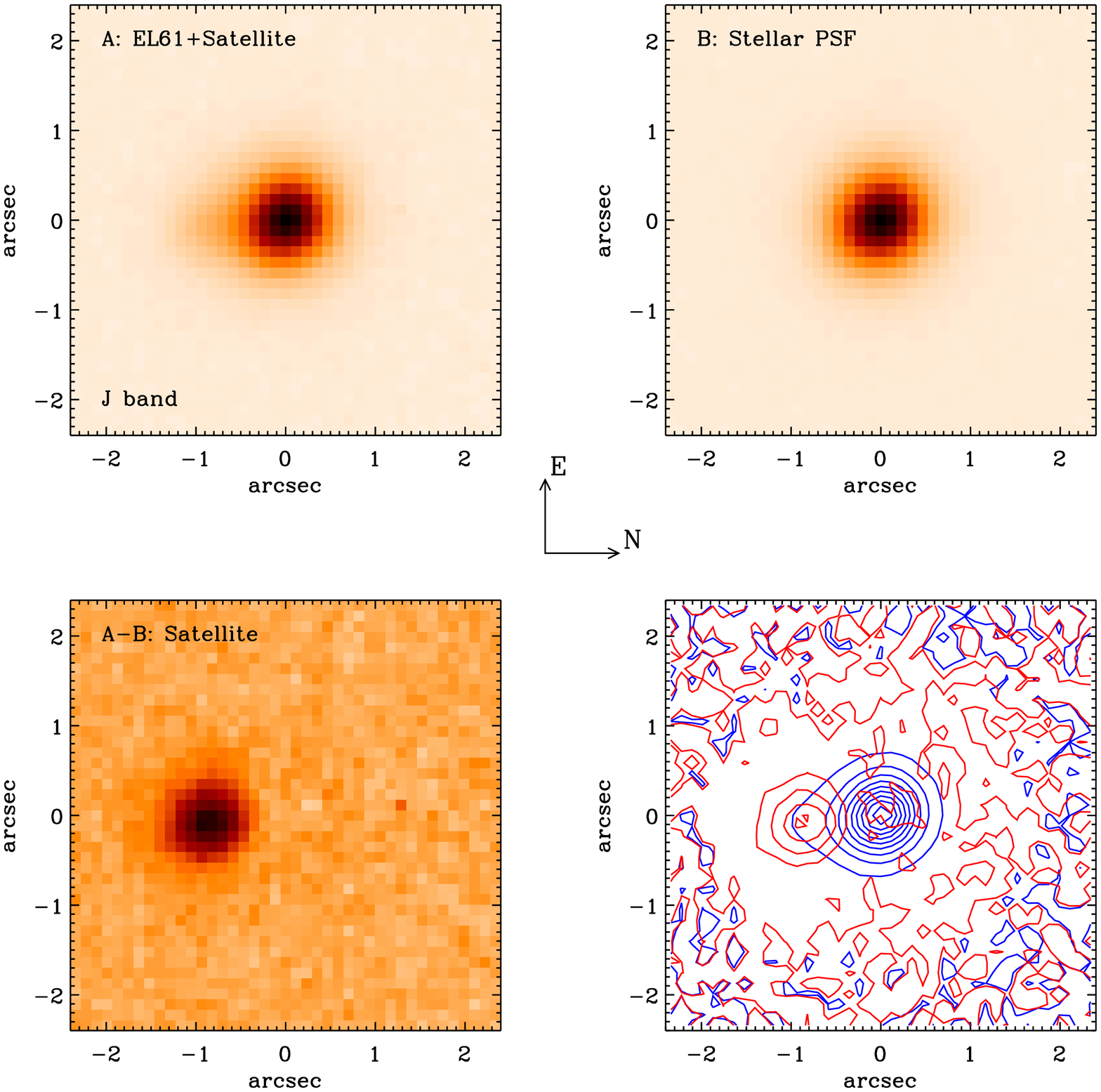}

   \caption {{\it Top left:} Stack of $\sim70$ $J$-band frames of Haumea.  The
satellite Hi'iaka is discernible to the left of Haumea. {\it Top right:}
Stellar PSF built from a stack of $\sim70$ $J$-band frames of a star in the
Haumea field.  {\it Bottom left:} Image of Hi'iaka obtained by subtracting the
top right image from the top left one.  {\it Bottom right:} Contours of Haumea
and Hi'iaka (blue) overplotted on contours of the satellite Hi'iaka alone (red)
to show the relative position.  } 

   \label{Fig.SatelliteJ}
   \end{figure*}
}

\def\FigSatelliteH{
   \begin{figure*}
   \centering
      \includegraphics[width=0.8\textwidth]{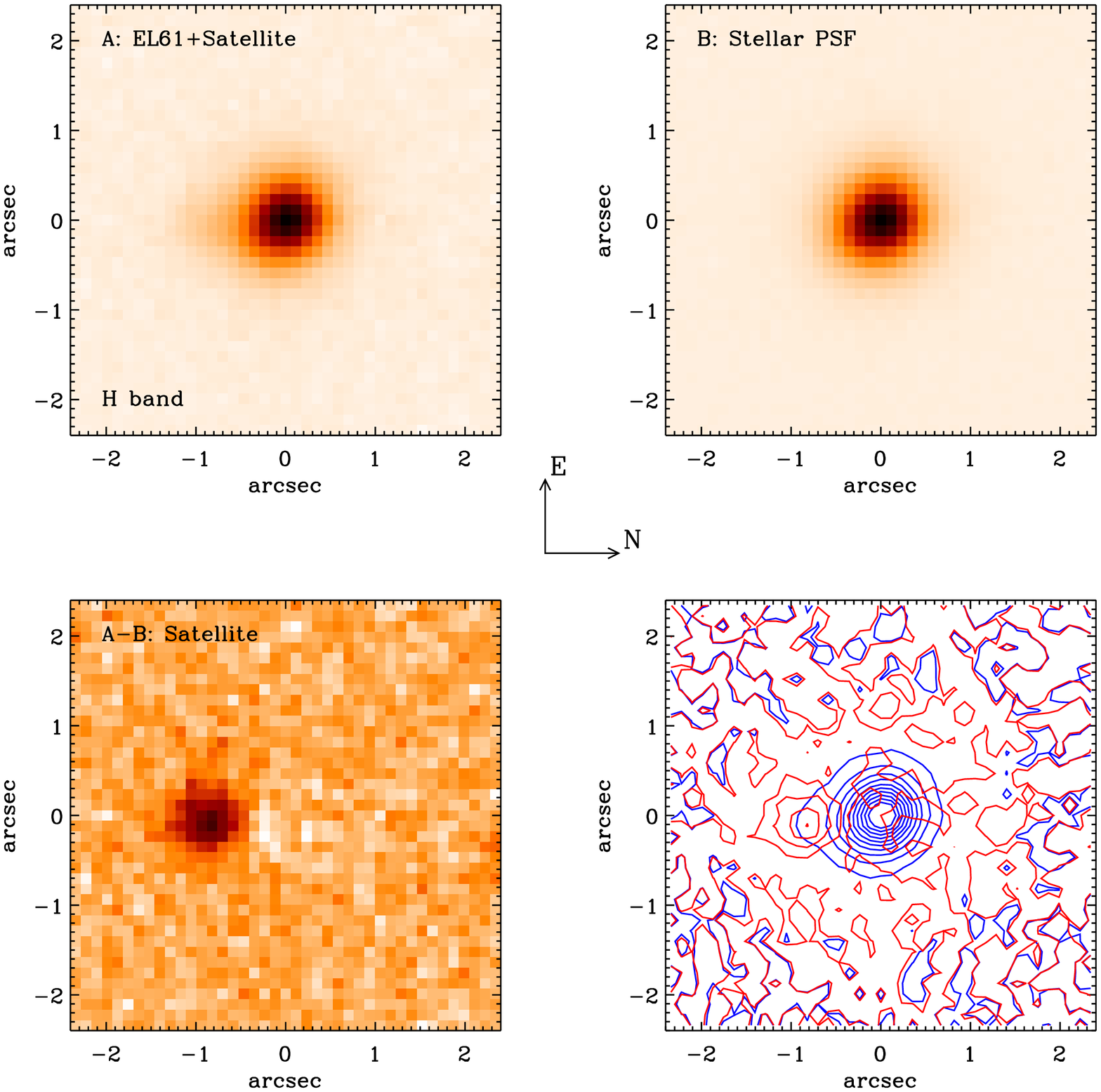}

   \caption {Same as Fig.~\ref{Fig.SatelliteJ} for $H$-band data.} 

   \label{Fig.SatelliteH}
   \end{figure*}
}

\begin{document}

\title{Time-Resolved Near-Infrared Photometry \\ of Extreme Kuiper Belt Object
Haumea}

\author{Pedro Lacerda}

\affil{Institute for Astronomy, University of Hawaii, 2680 Woodlawn
Drive, Honolulu, HI 96822}

\email{\myemail}

\begin{abstract}

We present time-resolved near-infrared ($J$ and $H$) photometry of the extreme
Kuiper belt object (136108) Haumea (formerly 2003$\,$EL$_{61}$) taken to
further investigate rotational variability of this object.  The new data show
that the near-infrared peak-to-peak photometric range is similar to the value
at visible wavelengths, $\Delta m_R$ = 0.30$\pm$0.02 mag.  Detailed analysis of
the new and previous data reveals subtle visible/near-infrared color variations
across the surface of Haumea.  The color variations are spatially correlated
with a previously identified surface region, redder in $B\!-\!R$ and darker
than the mean surface.  Our photometry indicates that the $J\!-\!H$ colors of
Haumea ($J\!-\!H=-0.057\pm0.016$ mag) and its brightest satellite Hi'iaka
($J\!-\!H=-0.399\pm0.034$ mag) are significantly ($>$9$\sigma$) different.  The
satellite Hi'iaka is unusually blue in $J\!-\!H$, consistent with strong 1.5
$\mu$m water-ice absorption.  The phase coefficient of Haumea in the $J$-band
is found to increase monotonically with wavelength in the range
$0.4<\lambda<1.3$.  We compare our findings with other Solar system objects and
discuss implications regarding the surface of Haumea.

\end{abstract}

\keywords{Kuiper belt --- methods: data analysis --- minor planets, asteroids
--- solar system: general --- techniques: photometric}

\section{Introduction}

Kuiper belt objects (KBOs) orbit the sun in the trans-Neptunian region of the
Solar system.  Mainly due to their large heliocentric distances and resulting
low temperatures, KBOs are amongst the least processed relics of the planetary
accretion disk and thus carry invaluable information about the physics and
chemistry of planet formation.  Moreover, as a surviving product of the debris
disk of the Sun, the Kuiper belt is a nearby analog to debris disks around
other stars and may provide useful insights into the study of the latter.

The known KBO population -- which currently amounts to over a thousand objects
-- provides several clues to the origin and evolution not only of the small
bodies but also of the planets.  One example is the outward migration of planet
Neptune, inferred from the need to explain the resonant structure of the KBO
population, namely the 3:2 resonant KBOs of which (134340) Pluto is a member
\citep{1995AJ....110..420Mal}.  Extreme, physically unusual objects are a
profitable source of interesting science as they often challenge existing
paradigms.  One such unusual object in the Kuiper belt is (136108) Haumea,
formerly known as 2003$\,$EL$_{61}$.  

Haumea is remarkable in many ways.  With approximate triaxial semi-axes
$1000\times800\times500$ km, it is one of the largest known KBOs.  Its
elongated shape is a consequence of the very rapid 3.9 h period rotation, and
those two properties combined can be used to infer Haumea's bulk density
($\rho\sim2500$ kg m$^{-3}$), assuming that the object's shape has relaxed to
hydrostatic equilibrium \citep{2006ApJ...639.1238R,2007AJ....133.1393L}.
Haumea's rapid rotation and the spectral and orbital similarity between this
object and a number of smaller KBOs, have led \citet{2007Natur.446..294Bro} to
suggest that an ancient shattering collision \citep[$>1$ Gyr
ago;][]{2007AJ....134.2160Rag} could explain both.  Haumea is one of the bluest
known KBOs, with $B\!-\!R=0.97\pm0.03$ mag
\citep{2006ApJ...639.1238R,2008AJ....135.1749Lac}, and it has an optical and
infrared spectrum consistent with a surface coated in almost pure water-ice
\citep{2007AJ....133..526T,2007ApJ...655.1172T}. This stands in contrast with
other large KBOs such as Pluto, Eris, and 2005 FY$_9$, which have methane rich
surfaces
\citep{1976Sci...194..835Cru,2005ApJ...635L..97Bro,2006A&A...445L..35Lic}.  Two
satellites have been detected in orbit around Haumea. The innermost, Namaka,
has an orbital period of $P_\mathrm{orb}\sim34$ days, an apparent orbital
semimajor axis $a\sim1\arcsec$, and a fractional optical brightness of
$f\sim1.5$\% with respect to Haumea.  The outermost, Hi'iaka has
$P_\mathrm{orb}\sim41$ days, $a\sim1.2\arcsec$, and $f\sim6$\%
\citep{2006ApJ...639L..43Bro}.

Time-resolved optical photometry of Haumea has revealed evidence for a
localized surface feature both redder and darker than the surrounding material
\citep{2008AJ....135.1749Lac}.  Although the existing data are unable to break
the degeneracy between the physical size and the color or albedo of this dark,
red spot (hereafter, DRS), the evidence points to it taking a large ($>20$\%)
fraction of the instantaneous cross section.  The composition of the DRS
remains unknown but its albedo and $B\!-\!R$ color are consistent with the
surfaces of Eris, 2005 FY$_9$, and Pluto's and Iapetus' brighter regions.
These observations motivated us to search for rotational modulation of the
water ice band strength that might be associated with the optically detected
DRS.

In this paper we provide further constraints on the surface properties of
Haumea.  We present time-resolved near-infrared ($J$ and $H$) data and
search for visible/near-infrared color variability.  We also constrain the
$J$-band phase function of Haumea and compare it to its optical counterparts.
Finally, we measure the $J\!-\!H$ color of Hi'iaka, the brightest satellite of
Haumea.

\section{Observations} \label{Sec.Observations}

\begin{deluxetable}{ll}[t]
  \tablecaption{Journal of Observations of Haumea. \label{Table.Journal}}
   \tablewidth{0pt}
   \startdata
\hline
  UT Date                       & 2008 Apr 14 \\
  Heliocentric Distance, $R$    & 51.116 AU\\
  Geocentric Distance, $\Delta$ & 50.240 AU\\
  Phase angle, $\alpha$         & 0.55\degr\\
  Weather                       & Photometric \\
  Telescope                     & 8.2 m Subaru \\
  Instrument                    & MOIRCS \\
  Pixel scale                   & 0.117\arcsec/pixel \\
  Seeing                        & 0.5\arcsec -- 0.8\arcsec \\
  Filters (Exp. Time)           & $J$ (30 s), $H$ (20 s) 
   \enddata
\end{deluxetable}

\begin{deluxetable*}{ccc|ccc}
  \tablecaption{$J$-band Photometry of Haumea. \label{Table.JData}}
   \tablewidth{0pt}
   \tablehead{
   \colhead{UT Date\tna} & \colhead{Julian Date\tna} & \colhead{$m_J$\tnb} &
   \colhead{UT Date\tna} & \colhead{Julian Date\tna} & \colhead{$m_J$\tnb}  
   }
   \startdata
2008 Apr 15.10985 & 2454571.609852 & 16.370$\pm$0.022 &  2008 Apr 15.26057 & 2454571.760570 & 16.416$\pm$0.022 \\
2008 Apr 15.11066 & 2454571.610665 & 16.352$\pm$0.022 &  2008 Apr 15.26138 & 2454571.761382 & 16.441$\pm$0.022 \\
2008 Apr 15.11175 & 2454571.611747 & 16.352$\pm$0.020 &  2008 Apr 15.26790 & 2454571.767900 & 16.411$\pm$0.022 \\
2008 Apr 15.11255 & 2454571.612546 & 16.338$\pm$0.024 &  2008 Apr 15.26872 & 2454571.768716 & 16.398$\pm$0.022 \\
2008 Apr 15.16761 & 2454571.667607 & 16.352$\pm$0.019 &  2008 Apr 15.26979 & 2454571.769791 & 16.404$\pm$0.024 \\
2008 Apr 15.16842 & 2454571.668420 & 16.334$\pm$0.019 &  2008 Apr 15.27060 & 2454571.770602 & 16.371$\pm$0.023 \\
2008 Apr 15.18420 & 2454571.684204 & 16.385$\pm$0.020 &  2008 Apr 15.27696 & 2454571.776959 & 16.309$\pm$0.022 \\
2008 Apr 15.18502 & 2454571.685020 & 16.367$\pm$0.019 &  2008 Apr 15.27777 & 2454571.777775 & 16.260$\pm$0.022 \\
2008 Apr 15.18610 & 2454571.686098 & 16.386$\pm$0.019 &  2008 Apr 15.27886 & 2454571.778859 & 16.305$\pm$0.022 \\
2008 Apr 15.18691 & 2454571.686910 & 16.390$\pm$0.019 &  2008 Apr 15.27967 & 2454571.779671 & 16.272$\pm$0.022 \\
2008 Apr 15.19373 & 2454571.693727 & 16.343$\pm$0.018 &  2008 Apr 15.28660 & 2454571.786601 & 16.189$\pm$0.026 \\
2008 Apr 15.19454 & 2454571.694539 & 16.340$\pm$0.019 &  2008 Apr 15.28741 & 2454571.787407 & 16.173$\pm$0.022 \\
2008 Apr 15.19562 & 2454571.695623 & 16.327$\pm$0.018 &  2008 Apr 15.28849 & 2454571.788485 & 16.200$\pm$0.024 \\
2008 Apr 15.19644 & 2454571.696435 & 16.332$\pm$0.018 &  2008 Apr 15.28930 & 2454571.789298 & 16.186$\pm$0.024 \\
2008 Apr 15.20279 & 2454571.702793 & 16.263$\pm$0.019 &  2008 Apr 15.29585 & 2454571.795847 & 16.158$\pm$0.021 \\
2008 Apr 15.20361 & 2454571.703606 & 16.234$\pm$0.019 &  2008 Apr 15.29666 & 2454571.796665 & 16.132$\pm$0.021 \\
2008 Apr 15.20468 & 2454571.704682 & 16.247$\pm$0.019 &  2008 Apr 15.29774 & 2454571.797743 & 16.112$\pm$0.022 \\
2008 Apr 15.20549 & 2454571.705494 & 16.237$\pm$0.019 &  2008 Apr 15.29856 & 2454571.798556 & 16.121$\pm$0.021 \\
2008 Apr 15.21199 & 2454571.711991 & 16.180$\pm$0.019 &  2008 Apr 15.30507 & 2454571.805070 & 16.130$\pm$0.020 \\
2008 Apr 15.21280 & 2454571.712804 & 16.174$\pm$0.019 &  2008 Apr 15.30588 & 2454571.805884 & 16.135$\pm$0.021 \\
2008 Apr 15.21390 & 2454571.713896 & 16.172$\pm$0.019 &  2008 Apr 15.30697 & 2454571.806969 & 16.162$\pm$0.021 \\
2008 Apr 15.21471 & 2454571.714708 & 16.186$\pm$0.019 &  2008 Apr 15.30778 & 2454571.807784 & 16.123$\pm$0.020 \\
2008 Apr 15.22109 & 2454571.721087 & 16.167$\pm$0.018 &  2008 Apr 15.31249 & 2454571.812494 & 16.191$\pm$0.020 \\
2008 Apr 15.22190 & 2454571.721900 & 16.182$\pm$0.017 &  2008 Apr 15.31330 & 2454571.813304 & 16.174$\pm$0.021 \\
2008 Apr 15.22297 & 2454571.722974 & 16.173$\pm$0.019 &  2008 Apr 15.31831 & 2454571.818312 & 16.212$\pm$0.021 \\
2008 Apr 15.22379 & 2454571.723786 & 16.198$\pm$0.019 &  2008 Apr 15.31913 & 2454571.819128 & 16.246$\pm$0.019 \\
2008 Apr 15.23047 & 2454571.730466 & 16.228$\pm$0.019 &  2008 Apr 15.32385 & 2454571.823849 & 16.286$\pm$0.019 \\
2008 Apr 15.23128 & 2454571.731280 & 16.252$\pm$0.019 &  2008 Apr 15.32466 & 2454571.824657 & 16.301$\pm$0.021 \\
2008 Apr 15.23236 & 2454571.732358 & 16.232$\pm$0.020 &  2008 Apr 15.32953 & 2454571.829527 & 16.330$\pm$0.025 \\
2008 Apr 15.23317 & 2454571.733171 & 16.261$\pm$0.019 &  2008 Apr 15.33034 & 2454571.830342 & 16.318$\pm$0.022 \\
2008 Apr 15.23952 & 2454571.739521 & 16.333$\pm$0.021 &  2008 Apr 15.33506 & 2454571.835055 & 16.351$\pm$0.022 \\
2008 Apr 15.24034 & 2454571.740339 & 16.336$\pm$0.022 &  2008 Apr 15.33586 & 2454571.835859 & 16.361$\pm$0.021 \\
2008 Apr 15.24140 & 2454571.741404 & 16.350$\pm$0.021 &  2008 Apr 15.34632 & 2454571.846321 & 16.375$\pm$0.022 \\
2008 Apr 15.24222 & 2454571.742216 & 16.345$\pm$0.020 &  2008 Apr 15.34713 & 2454571.847132 & 16.377$\pm$0.020 \\
2008 Apr 15.24959 & 2454571.749588 & 16.392$\pm$0.020 &  2008 Apr 15.34822 & 2454571.848217 & 16.382$\pm$0.021 \\
2008 Apr 15.25039 & 2454571.750392 & 16.425$\pm$0.020 &  2008 Apr 15.34911 & 2454571.849106 & 16.352$\pm$0.020 \\
2008 Apr 15.25147 & 2454571.751474 & 16.432$\pm$0.020 &  2008 Apr 15.35012 & 2454571.850125 & 16.347$\pm$0.020 \\
2008 Apr 15.25229 & 2454571.752288 & 16.415$\pm$0.021 &  2008 Apr 15.35094 & 2454571.850936 & 16.366$\pm$0.020 \\
2008 Apr 15.25867 & 2454571.758673 & 16.420$\pm$0.022 &  2008 Apr 15.35202 & 2454571.852017 & 16.366$\pm$0.020 \\
2008 Apr 15.25949 & 2454571.759489 & 16.409$\pm$0.024 &  2008 Apr 15.35283 & 2454571.852828 & 16.347$\pm$0.020 \\
   \enddata
  \tablenotetext{a}{Dates in Haumea's reference frame;}
  \tablenotetext{b}{Apparent magnitude.}
\end{deluxetable*}

\begin{deluxetable*}{ccc|ccc}
  \tablecaption{$H$-band Photometry of Haumea. \label{Table.HData}}
   \tablewidth{0pt}
   \tablehead{
   \colhead{UT Date\tna} & \colhead{Julian Date\tna} & \colhead{$m_H$\tnb} &
   \colhead{UT Date\tna} & \colhead{Julian Date\tna} & \colhead{$m_H$\tnb}  
   }
   \startdata
2008 Apr 15.11484 & 2454571.614839 & 16.370$\pm$0.026 &  2008 Apr 15.25430 & 2454571.754303 & 16.505$\pm$0.024 \\
2008 Apr 15.11553 & 2454571.615528 & 16.396$\pm$0.026 &  2008 Apr 15.25501 & 2454571.755009 & 16.493$\pm$0.022 \\
2008 Apr 15.11648 & 2454571.616483 & 16.355$\pm$0.024 &  2008 Apr 15.25596 & 2454571.755956 & 16.489$\pm$0.022 \\
2008 Apr 15.11719 & 2454571.617187 & 16.382$\pm$0.025 &  2008 Apr 15.25665 & 2454571.756649 & 16.474$\pm$0.022 \\
2008 Apr 15.17144 & 2454571.671445 & 16.449$\pm$0.021 &  2008 Apr 15.26340 & 2454571.763400 & 16.478$\pm$0.022 \\
2008 Apr 15.17213 & 2454571.672129 & 16.423$\pm$0.020 &  2008 Apr 15.26409 & 2454571.764093 & 16.490$\pm$0.026 \\
2008 Apr 15.17311 & 2454571.673107 & 16.426$\pm$0.021 &  2008 Apr 15.26505 & 2454571.765051 & 16.527$\pm$0.025 \\
2008 Apr 15.17380 & 2454571.673799 & 16.424$\pm$0.021 &  2008 Apr 15.26574 & 2454571.765742 & 16.500$\pm$0.026 \\
2008 Apr 15.18920 & 2454571.689196 & 16.430$\pm$0.021 &  2008 Apr 15.27262 & 2454571.772622 & 16.388$\pm$0.025 \\
2008 Apr 15.18989 & 2454571.689885 & 16.402$\pm$0.021 &  2008 Apr 15.27332 & 2454571.773318 & 16.455$\pm$0.028 \\
2008 Apr 15.19086 & 2454571.690862 & 16.414$\pm$0.022 &  2008 Apr 15.27427 & 2454571.774272 & 16.432$\pm$0.031 \\
2008 Apr 15.19155 & 2454571.691552 & 16.413$\pm$0.022 &  2008 Apr 15.27496 & 2454571.774963 & 16.401$\pm$0.027 \\
2008 Apr 15.19847 & 2454571.698466 & 16.349$\pm$0.019 &  2008 Apr 15.28168 & 2454571.781679 & 16.273$\pm$0.027 \\
2008 Apr 15.19915 & 2454571.699147 & 16.340$\pm$0.020 &  2008 Apr 15.28237 & 2454571.782369 & 16.333$\pm$0.033 \\
2008 Apr 15.20011 & 2454571.700109 & 16.335$\pm$0.019 &  2008 Apr 15.28332 & 2454571.783321 & 16.302$\pm$0.031 \\
2008 Apr 15.20079 & 2454571.700787 & 16.330$\pm$0.017 &  2008 Apr 15.28401 & 2454571.784011 & 16.290$\pm$0.034 \\
2008 Apr 15.20749 & 2454571.707494 & 16.261$\pm$0.019 &  2008 Apr 15.29131 & 2454571.791314 & 16.249$\pm$0.024 \\
2008 Apr 15.20819 & 2454571.708186 & 16.268$\pm$0.020 &  2008 Apr 15.29201 & 2454571.792008 & 16.210$\pm$0.023 \\
2008 Apr 15.20914 & 2454571.709144 & 16.233$\pm$0.020 &  2008 Apr 15.29298 & 2454571.792983 & 16.188$\pm$0.021 \\
2008 Apr 15.20984 & 2454571.709836 & 16.225$\pm$0.020 &  2008 Apr 15.29367 & 2454571.793674 & 16.165$\pm$0.022 \\
2008 Apr 15.21673 & 2454571.716730 & 16.238$\pm$0.018 &  2008 Apr 15.30058 & 2454571.800584 & 16.186$\pm$0.020 \\
2008 Apr 15.21742 & 2454571.717424 & 16.197$\pm$0.018 &  2008 Apr 15.30126 & 2454571.801264 & 16.192$\pm$0.019 \\
2008 Apr 15.21838 & 2454571.718383 & 16.235$\pm$0.019 &  2008 Apr 15.30222 & 2454571.802219 & 16.181$\pm$0.020 \\
2008 Apr 15.21907 & 2454571.719073 & 16.208$\pm$0.019 &  2008 Apr 15.30291 & 2454571.802912 & 16.179$\pm$0.021 \\
2008 Apr 15.22582 & 2454571.725818 & 16.246$\pm$0.019 &  2008 Apr 15.30979 & 2454571.809794 & 16.213$\pm$0.023 \\
2008 Apr 15.22651 & 2454571.726511 & 16.245$\pm$0.020 &  2008 Apr 15.31049 & 2454571.810488 & 16.220$\pm$0.023 \\
2008 Apr 15.22747 & 2454571.727468 & 16.261$\pm$0.019 &  2008 Apr 15.31532 & 2454571.815318 & 16.253$\pm$0.024 \\
2008 Apr 15.22816 & 2454571.728161 & 16.255$\pm$0.019 &  2008 Apr 15.31601 & 2454571.816013 & 16.253$\pm$0.023 \\
2008 Apr 15.23518 & 2454571.735181 & 16.349$\pm$0.021 &  2008 Apr 15.32114 & 2454571.821135 & 16.333$\pm$0.024 \\
2008 Apr 15.23588 & 2454571.735878 & 16.344$\pm$0.020 &  2008 Apr 15.32183 & 2454571.821831 & 16.287$\pm$0.024 \\
2008 Apr 15.23683 & 2454571.736832 & 16.369$\pm$0.023 &  2008 Apr 15.32668 & 2454571.826679 & 16.383$\pm$0.026 \\
2008 Apr 15.23751 & 2454571.737510 & 16.343$\pm$0.021 &  2008 Apr 15.32736 & 2454571.827364 & 16.389$\pm$0.025 \\
2008 Apr 15.24423 & 2454571.744229 & 16.394$\pm$0.023 &  2008 Apr 15.33236 & 2454571.832360 & 16.457$\pm$0.028 \\
2008 Apr 15.24492 & 2454571.744916 & 16.408$\pm$0.024 &  2008 Apr 15.33305 & 2454571.833051 & 16.393$\pm$0.026 \\
2008 Apr 15.24587 & 2454571.745869 & 16.428$\pm$0.023 &  2008 Apr 15.33789 & 2454571.837886 & 16.443$\pm$0.028 \\
2008 Apr 15.24656 & 2454571.746561 & 16.459$\pm$0.023 &  2008 Apr 15.33861 & 2454571.838607 & 16.439$\pm$0.027 \\
   \enddata
  \tablenotetext{a}{Dates in Haumea's reference frame;}
  \tablenotetext{b}{Apparent magnitude.}
\end{deluxetable*}

\begin{deluxetable}{ccc}
  \tablecaption{Binned $J$-band Data. \label{Table.MeanJData}}
   \tablewidth{0pt}
   \tablehead{
   \colhead{Mean UT Date\tna} & \colhead{Mean Julian Date\tna} & \colhead{Mean $m_J$\tnb}  
   }
   \startdata
2008 Apr 15.11120 & 2454571.611202 & 16.353$\pm$0.013 \\
2008 Apr 15.16801 & 2454571.668013 & 16.343$\pm$0.013 \\
2008 Apr 15.18556 & 2454571.685558 & 16.382$\pm$0.010 \\
2008 Apr 15.19508 & 2454571.695081 & 16.336$\pm$0.009 \\
2008 Apr 15.20414 & 2454571.704144 & 16.245$\pm$0.010 \\
2008 Apr 15.21335 & 2454571.713350 & 16.178$\pm$0.009 \\
2008 Apr 15.22244 & 2454571.722437 & 16.180$\pm$0.010 \\
2008 Apr 15.23182 & 2454571.731819 & 16.243$\pm$0.011 \\
2008 Apr 15.24087 & 2454571.740870 & 16.341$\pm$0.010 \\
2008 Apr 15.25094 & 2454571.750936 & 16.416$\pm$0.012 \\
2008 Apr 15.26003 & 2454571.760028 & 16.422$\pm$0.012 \\
2008 Apr 15.26925 & 2454571.769252 & 16.396$\pm$0.013 \\
2008 Apr 15.27832 & 2454571.778316 & 16.287$\pm$0.015 \\
2008 Apr 15.28795 & 2454571.787948 & 16.187$\pm$0.012 \\
2008 Apr 15.29720 & 2454571.797203 & 16.131$\pm$0.013 \\
2008 Apr 15.30643 & 2454571.806427 & 16.138$\pm$0.012 \\
2008 Apr 15.31290 & 2454571.812899 & 16.183$\pm$0.014 \\
2008 Apr 15.31872 & 2454571.818720 & 16.229$\pm$0.018 \\
2008 Apr 15.32425 & 2454571.824253 & 16.294$\pm$0.013 \\
2008 Apr 15.32993 & 2454571.829934 & 16.324$\pm$0.015 \\
2008 Apr 15.33546 & 2454571.835457 & 16.356$\pm$0.013 \\
2008 Apr 15.34769 & 2454571.847694 & 16.372$\pm$0.011 \\
2008 Apr 15.35148 & 2454571.851476 & 16.357$\pm$0.010 \\
   \enddata
  \tablenotetext{a}{Dates in Haumea's reference frame;}
  \tablenotetext{b}{Mean apparent magnitude.}
\end{deluxetable}

\begin{deluxetable}{ccc}
  \tablecaption{Binned $H$-band Data. \label{Table.MeanHData}}
   \tablewidth{0pt}
   \tablehead{
   \colhead{Mean UT Date\tna} & \colhead{Mean Julian Date\tna} & \colhead{Mean $m_H$\tnb}  
   }
   \startdata
2008 Apr 15.11601 & 2454571.616009 & 16.376$\pm$0.015 \\
2008 Apr 15.17262 & 2454571.672620 & 16.431$\pm$0.011 \\
2008 Apr 15.19037 & 2454571.690374 & 16.415$\pm$0.011 \\
2008 Apr 15.19963 & 2454571.699627 & 16.339$\pm$0.009 \\
2008 Apr 15.20866 & 2454571.708665 & 16.247$\pm$0.013 \\
2008 Apr 15.21790 & 2454571.717903 & 16.220$\pm$0.012 \\
2008 Apr 15.22699 & 2454571.726990 & 16.252$\pm$0.009 \\
2008 Apr 15.23635 & 2454571.736350 & 16.352$\pm$0.011 \\
2008 Apr 15.24539 & 2454571.745394 & 16.423$\pm$0.016 \\
2008 Apr 15.25548 & 2454571.755479 & 16.491$\pm$0.012 \\
2008 Apr 15.26457 & 2454571.764572 & 16.499$\pm$0.014 \\
2008 Apr 15.27379 & 2454571.773793 & 16.419$\pm$0.018 \\
2008 Apr 15.28284 & 2454571.782845 & 16.300$\pm$0.018 \\
2008 Apr 15.29249 & 2454571.792495 & 16.203$\pm$0.019 \\
2008 Apr 15.30174 & 2454571.801745 & 16.185$\pm$0.009 \\
2008 Apr 15.31014 & 2454571.810141 & 16.217$\pm$0.014 \\
2008 Apr 15.31567 & 2454571.815666 & 16.253$\pm$0.013 \\
2008 Apr 15.32148 & 2454571.821483 & 16.310$\pm$0.023 \\
2008 Apr 15.32702 & 2454571.827022 & 16.386$\pm$0.015 \\
2008 Apr 15.33271 & 2454571.832706 & 16.425$\pm$0.030 \\
2008 Apr 15.33825 & 2454571.838247 & 16.441$\pm$0.016 \\
   \enddata
  \tablenotetext{a}{Dates in Haumea's reference frame;}
  \tablenotetext{b}{Mean apparent magnitude.}
\end{deluxetable}

\begin{deluxetable*}{cccccc}
  \tablecaption{Mean magnitudes and colors. \label{Table.Colors}}
   \tablewidth{0pt}
   \tablehead{
   \colhead{Object}& \colhead{$m_J$\tna} & \colhead{$m_H$\tnb} & \colhead{$B-R$} & \colhead{$R-J$} & \colhead{$J-H$} 
   }
   \startdata
   Haumea         & $16.29\pm0.09$ & $16.34\pm0.10$ & $0.972\pm0.014$ & $0.885\pm0.012$ & $-0.057\pm0.016$ \\
   Satellite      & $18.92\pm0.09$ & $19.32\pm0.10$ & $\dots$         & $\dots$         & $-0.399\pm0.034$ \\
   \enddata
  \tablenotetext{a}{Mean apparent $J$ magnitude on 2008 Apr 15 UT;}
  \tablenotetext{b}{Mean apparent $H$ magnitude on 2008 Apr 15 UT;}
\end{deluxetable*}

Near-infrared observations were taken using the 8.2-m diameter Subaru telescope
atop Mauna Kea, Hawaii.  We used the Multi-Object Infrared Camera and
Spectrograph \citep[MOIRCS;][]{2003SPIE.4841.1625T} which is mounted at the
f/12.2 Cassegrain focus.  MOIRCS accomodates two 2048$\times$2048 pixel HgCdTe
(HAWAII-2) arrays, with each pixel projecting onto a square 0.117\arcsec\ on a
side in the sky.  Observations were obtained through broadband $J$
($\lambda_\mathrm{c}=1.26$ \mum, $\Delta\lambda=0.17$ \mum) and $H$
($\lambda_\mathrm{c}=1.64$ \mum, $\Delta\lambda=0.28$ \mum) filters.  The data
were instrumentally calibrated using dark frames and dome flat-field images
obtained immediately before and after the night of observation.  Because of
technical difficulties with detector 1, we used detector 2 for all our science
and calibration frames.  Science images were obtained in sets of two dithered
positions 15\arcsec\ apart, which were later mutually subtracted to remove the
infrared background flux.  

The night of 2008 April 15 UT was photometric, allowing us to absolutely
calibrate the data using observations of standard star FS33 from the UKIRT
Faint Standards catalog \citep{2001MNRAS.325..563Haw}.  The Haumea flux through
each filter was measured using circular aperture photometry relative to a field
star, while a second field star was used to verify the constancy of the first.
The dispersion in the star-to-star relative photometry indicates a mean
$1\sigma$ uncertainty of $\pm0.015$ magnitude in $J$ and $\pm0.023$ magnitude
in $H$.  The field star was calibrated to the standard star FS33 at airmass
$1.02$, just short of the telescope's Alt-Az elevation limit.  From scatter in
the standard star photometry we estimate a systematic uncertainty in the
absolute calibration of $0.04$ magnitudes in $J$ and $0.02$ magnitudes in $H$.
A brief journal of observations can be found in Table~\ref{Table.Journal}.  The
final calibrated broadband photometric measurements are listed in
Tables~\ref{Table.JData} and \ref{Table.HData}.

We generally obtained two consecutive sets of two dithered images in each
filter before switching filters (i.e. $JJ$-$JJ$-$HH$-$HH$-$\dots$).  This
results in sets of four data points all within 3 to 4 minutes of each other.
To reduce the scatter in the lightcurves we binned each of these sets of
consecutive measurements into a single data point, with each binned point
obtained by averaging the times and magnitudes of the set. The error bar on
each binned point includes the error on the mean magnitude and the average
uncertainty of the unbinned measurements, added in quadrature.  The binned
measurements are listed in Tables~\ref{Table.MeanJData} and
\ref{Table.MeanHData}.

\vfil

\section{Results and Discussion}

\subsection{Color Versus Rotation}

\FigLightcurve

\FigLightcurveBinned

The new data record just over one full rotation of Haumea.
Figure~\ref{Fig.Lightcurve} combines previously published $B$ and $R$ data
\citep{2008AJ....135.1749Lac} with the new $J$ and $H$ data and shows that all
four filters exhibit very similar variability with a combined total range
$\Delta m=0.30\pm0.02$ mag.  As described in \S\ref{Sec.Observations}, to
improve the signal-to-noise ratios of the $J$ and $H$ data, we binned sets of
measurements taken back-to-back (usually sets of four); the resulting
lightcurve is shown in Fig.~\ref{Fig.LightcurveBinned}.  There, the previously
identified dark, red spot \citep[DRS;][]{2008AJ....135.1749Lac} on the surface
of Haumea is clearly apparent at rotational phases close to $\phi=0.8$ in the
$B$ and $R$ curves.  The near-infrared data generally follow the $R$-band data
but show a slight visible/near-infrared reddening which coincides with the DRS.

The differences between the individual lightcurves in
Figs.~\ref{Fig.Lightcurve} and \ref{Fig.LightcurveBinned} are small.  To
highlight color variations on Haumea, we plot in Fig.~\ref{Fig.ColorCurves} all
possible combinations of visible-to-near-infrared color curves. The curves are
calculated by interpolating the better-sampled $B$ and $R$ data to the binned
$J$ and $H$ rotational phases (Fig.~\ref{Fig.LightcurveBinned}) and
subtracting.  The error bars are dominated by the uncertainties in the
near-infrared measurements, which are added quadratically to the mean $B$ or
$R$ errors.  When taken separately, the color curves in
Fig.~\ref{Fig.ColorCurves} appear to differ only marginally from a rotationally
constant value.  However, the color $B$$-$$H$, and arguably $B$$-$$J$,
$R$$-$$H$, and $J$$-$$H$, show visible reddening humps for rotational phases
close to where the DRS was found to lie ($\phi\sim0.8$).  To locate and
quantify color variability features in the curves in Fig.~\ref{Fig.ColorCurves}
we employ a running Gaussian probability test.  In this test we consider a
moving rotational phase window and calculate the quantity \begin{equation} G=
\frac{\sum_{i=1}^{N}(c_i - c_0)/e_i}{\sqrt{N}} \label{Eqn.G} \end{equation} for
the points that fall within the window.  In Eq.~(\ref{Eqn.G}), $N$ is the
number of points within the window, $c_i$ and $e_i$ are the color values and
respective error bars of those points, and $c_0$ is the median color of all
points (dotted horizontal lines in Fig.~\ref{Fig.ColorCurves}).  We then move
the window along each color curve in rotational phase steps of 0.05 to obtain a
running-$G$ value.  Equation (\ref{Eqn.G}) represents a Gaussian deviate with
zero mean and unity standard deviation and can thus be converted to a Gaussian
probability, $p(G)$, assuming that the points are normally distributed around
the median.  The probability $p(G)$ is sensitive to unlikely sequences of
deviant points, all on one side of the median.  Figure~\ref{Fig.GaussianProb}
shows the test results for each of the color curves using a window size
$\Delta\phi=0.25$ (see discussion below).  The Figure shows that the $B$$-$$H$
curve has a significant ($\sim$4$\sigma$) non-random feature close to
$\phi=0.8$.  The test also detects weaker (2.8$\sigma$ and 2.5$\sigma$)
features in the $B$$-$$J$ and $R$$-$$H$ curves close to $\phi=0.8$.

\FigColorcurves

\FigGaussianProb

\FigGaussianProbVsWindow

The size of the rotational phase window $\Delta\phi$ is physically motivated by
the fraction of the surface of Haumea that is visible at any given instant.  In
that sense, it should not be larger than $\Delta\phi=0.5$.  Moreover, although
half the surface is visible, projection effects in the limb region will make
the {\em effective} visible area smaller, by possibly another factor 2.  In
Fig.~\ref{Fig.GaussianProbVsWindow} we illustrate the effect of the window size
by replotting the running $p(G)$ for color curve $B$$-$$H$ using four window
sizes, $\Delta\phi=0.20$, 0.25, 0.33, and 0.50.  As expected, $p(G)$ does not
differ much for windows $0.20\leq\Delta\phi\leq0.33$.  For the largest window
size $\Delta\phi=0.50$ the probability begins to appear dilluted, but even then
the test succeeds in locating the feature at $\phi=0.8$.

The near-infrared measurements presented here are considerably less numerous
than the the optical data that were used to identify the DRS in $B$$-$$R$
\citep{2008AJ....135.1749Lac}.  Also, the $J$ and $H$ measurements may show
systematic correlations because they were measured on the same night using the
same telescope.  Nevertheless, the observed changes in $B$ and $R$ relative to
$J$ and $H$ do not suffer from this effect and are likely to be real.  A
visible/near-infrared reddening was already observed in our 2007 data (see
$J_{2007}$ points in Fig.~\ref{Fig.Lightcurve}) adding confidence to our
conclusions.  The results presented above suggest that the region close to the
DRS is also spectrally anomalous in the visible-to-near-infrared wavelength
range with respect to the average surface of Haumea.    

\FigSpectrumVsPhase

In Fig.~\ref{Fig.SpectrumVsPhase} we combine our four-band data to produce
reflectivity vs.\ wavelength curves at different rotational phases.  We focus
on the DRS region and plot curves at $\phi=0.3$ and $\phi=0.4$ as illustrative
of the mean Haumea surface.  We employed interpolation to calculate color
indices at the given rotational phases, which were subsequently converted to
reflectivities relative to the $R$-band. To enhance the subtle differences with
rotation we normalize all curves by that at $\phi=0.4$; an inset in
Fig.~\ref{Fig.SpectrumVsPhase} shows the reflectivities before normalization.
Figure~\ref{Fig.SpectrumVsPhase} shows that relative to the majority of the
surface of Haumea the region near $\phi=0.7$ displays an enhanced $H$-band
reflectivity which, close to $\phi=0.8$, is accompanied by a depressed $B$-band
reflectivity.  Close to $\phi=0.9$, $B$ remains depressed, while $H$ is
restored to average values.  We note that this variation is consistent with the
results shown in Fig.~\ref{Fig.GaussianProb}.  To summarize, the DRS region is
both fainter in $B$ and brighter in $H$ than the rest of Haumea.

The presence of a blue absorber on the DRS could explain the fainter $B$
reflectance.  A recent $U$- and $B$-band photometric study of KBOs
\citep{2007AJ....134.2046Jew} suggests that objects in the classical population
(objects in quasi-circular orbits between the 2:1 and the 3:2 mean-motion
resonance with Neptune) lack significant blue absorption.  As discussed by
those authors, $B$-band absorption in main-belt asteroids is generally linked
with the presence of phyllosilicates and other hydrated minerals \citep[see
also][]{1978SSRv...21..555Gaf,1985Icar...63..183Fei} and is a characteristic
feature in the spectra of C and G-type asteroids \citep{1989aste.conf..298Tho}.
How likely is it that Haumea has hydrated minerals on its surface?  Although
Haumea is located in the classical population as defined above, it is atypical
in its water-ice dominated surface spectrum and its high bulk density (2.5
times water).  These two properties suggest a differentiated body with
significant rocky content.  Furthermore, KBOs (mainly the larger ones) have
possibly sustained liquid water in their interiors due to radiogenic heating
\citep{2006Icar..183..283Mer}.  It would therefore be unremarkable to find
trace amounts of hydrated minerals on Haumea.  In fact, fits to the
near-infrared spectrum of Haumea are slightly improved by the addition of
phyllosilicates such as kaolinite and montmorillonite
\citep{2007ApJ...655.1172T}.  Ultraviolet spectra of Haumea at different
rotational phases are needed to further constrain the character of the blue
absorption close to the DRS.

Our $J$ and $H$ photometry suggests that the 1.5 $\mu$m water-ice band is
weaker (less deep) close to the DRS.  In contrast,
\citet{2008AJ....135.1749Lac} found marginal evidence that the 1.5 \mum\ band
is {\em deeper} close to the DRS.  One difference between the two measurements
is that while here we use $J$ vs.\ $H$, \citet{2008AJ....135.1749Lac} used $J$
vs.\ the ``CH4$_s$'' filter to assess possible variations in the water-ice
band.  The latter filter has a bandpass (center 1.60 \mum, FWHM 0.11 \mum)
between the 1.5 \mum\ and the 1.65 \mum\ band diagnostic of crystalline
water-ice and is thus affected by the degree of crystallinity of the ice.  The
two measurements can be reconciled if the DRS material has an overall less deep
1.5 \mum\ water-ice band but a larger relative abundance of crystalline water
ice.  We simulated this scenario using synthetic reflectance spectra
[calculated using published optical constants for crystalline water ice
\citep{1998JGR...10325809Gru} and a Hapke model with the best fit parameters
for Haumea \citep{2007ApJ...655.1172T}].  By convolving two model spectra, one
for $T=30$ K and one for $T=140$ K (to simulate a weaker crystalline band),
with the $J$, $H$, and CH4$_s$ bandpasses we found an effect similar to what is
observed.  The 30 K spectrum, taken to represent the DRS material, shows a
$\sim$3\% higher $H$-to-$J$ flux ratio, but a $\sim$4\% lower CH$_s$-to-$J$
flux ratio than the 140~K spectrum.  However, this possibility is not unique
and time-resolved $H$-band spectra are required to test this scenario.

\subsection{Phase Function}

Atmosphereless solar system bodies exhibit a linear increase in brightness with
decreasing phase angle.  At small angles ($\alpha<0.01$ to 1 deg), this phase
function becomes non-linear causing a sharp magnitude peak.  The main physical
mechanisms thought to be responsible for this opposition brightening effect are
shadowing and coherent backscattering.  In simple terms, shadowing occurs
because although a photon can always scatter back in the direction from which
it hit the surface, other directions may be blocked.  The implication is that
back-illuminated ($\alpha\sim0$\degr) objects do not shadow their own surfaces
and appear brighter.  The brightening due to coherent backscattering results
from the constructive interference of photons that scatter in the backwards
direction from pairs of surface particles \citep[or of features within a
particle;][]{2000Icar..147..545Nel}.  The constructive interfence decreases
rapidly with increasing phase angle.  Generally, it is believed that shadowing
regulates the decrease in brightness with phase angle from a few up to tens of
degrees, while coherent backscattering mainly produces the near-zero phase
angle spike \citep{2007PASP..119..623Fre}.

\FigPhaseCoefficients

\FigDBetaDLambdaVsAlbedo

The relative importance of shadowing and coherent backscattering on a given
surface is difficult to assess.  An early prediction by
\citet{1993tres.book.....Hap} was that the angular width of the exponential
brightness peak should vary linearly with wavelength in the case of coherent
backscattering, given its interference nature.  Shadowing, on the other hand,
should be acromatic.  However, more recent work has shown that the wavelength
dependence of coherent backscattering can be very weak
\citep{2000Icar..147..545Nel}.  Thus, while a strong wavelength dependence is
usually attributable to coherent backscattering, a weak wavelength dependence
may be explained by either mechanism.  It was also expected that higher albedo
surfaces should be less affected by shadowing because more multiply scattered
photons will reach the observer even from shadowed surface regions
\citep{1998Icar..131..223Nel}.  Coherent backscattering is a
multiple-scattering process that thrives on highly reflective surfaces.
Subsequent studies have shown that low albedo surfaces can also display strong
coherent backscattering at very low phase angles \citep{1998Icar..133...89Hap}.
Finally, while both effects cause a brightening towards opposition, only
coherent backscattering has an effect on the polarization properties of light
scattered from those objects: it favors the electric field component parallel
to the scattering plane and thus gives rise to partially linearly polarized
light \citep{2004A&A...415L..21Boe,2006A&A...450.1239Bag}.  With the currently
available instruments, useful polarization measurements can only be obtained on
the brightest KBOs.

Our July 2007 $J$-band lightcurve data \citep{2008AJ....135.1749Lac} and the
April 2008 data presented here together show that Haumea appears brighter in
the latter dataset.  After calculating the time-medianed apparent magnitudes,
and correcting them to unit helio- and geocentric distances ($R$ and $\Delta$)
using $m_J(1,1,\alpha) = m_J - 5\log(R\Delta)$, a difference of 0.096
magnitudes remains between the datasets, which we attribute to the different
phase angles ($\alpha_{2007}=1.13$ deg and $\alpha_{2008}=0.55$ deg) at the two
epochs.  Assuming a phase function of the form $m_J(\alpha)=m_J(1,1,0)+\beta_J
\alpha$, we derive a slope $\beta_J=0.16\pm0.04$ mag deg$^{-1}$.  The
uncertainty in $\beta_J$ was calculated by interpolating the 2008 $J$
lightcurve to the rotational phases of the 2007 $J$ lightcurve (both corrected
to $R=\Delta=1$ AU) and calculating the standard deviation of the difference.

In Fig.~\ref{Fig.PhaseCoefficients} we plot the phase coefficient $\beta$
versus the central wavelengths of the filters $B$, $V$, $I$, and $J$.  The
visible $\beta_{B,V,I}$ values are taken from \citet{2006ApJ...639.1238R} where
they are measured in the phase range $0.5<\alpha<1.0$ deg.  A linear fit to the
relation is overplotted as a dotted line showing that $\beta_\lambda$ increases
with $\lambda$.  
The fit has a slope $\Delta \beta/\Delta \lambda = 0.09\pm0.03$ mag deg$^{-1}$
$\mu$m$^{-1}$ and a 1 $\mu$m value $\beta_{1\mu\mathrm{m}} = 0.14\pm0.02$ mag
deg$^{-1}$.  Using a $\chi^2$ test we are only able to reject a constant
$\beta_\lambda$ at the 2$\sigma$ level.  However, the monotonic increase of
$\beta_\lambda$ with wavelength plus the high albedo
\citep[$p_V>0.6$,][]{2006ApJ...639.1238R} of Haumea both suggest that coherent
backscattering is dominant in the range of phase angles observed.

Small icy bodies in the solar system show very diverse phase function vs.\
wavelength behaviors \citep{2007AJ....133...26R}.  Most KBOs for which the
phase curve has been measured in more than one band show steeper slopes towards
longer wavelengths.  We show here that in the case of Haumea this behavior
extends into the near-infrared.  The Centaurs \citep{2007AJ....133...26R} and
the Uranian satellites \citep{2001Icar..151...51Kar} show little variation in
the phase curve with wavelength for $\alpha < 3$ to 6 deg, suggesting that
shadowing is dominant over coherent backscattering.  Other objects show
opposite or even non-monotonic relations between phase function slope and
wavelength.  For instance, some type of terrain on the Jovian moon Europa have
phase functions that vary non-monotonically with wavelength
\citep{1998Icar..135...41Hel}, while Pluto's phase function has a weak
wavelength dependence opposite to that seen in Haumea, from
$\beta_B=0.037\pm0.001$ mag deg$^{-1}$ to $\beta_R=0.032\pm0.001$ mag
deg$^{-1}$ \citep{2003Icar..162..171Bur}.

In Fig.~\ref{Fig.DBetaDLambdaVsAlbedo} we plot the slope of the $\beta_\lambda$
vs.\ $\lambda$ relation against approximate geometric albedo for a number of
KBOs and Centaurs.  The values $\Delta\beta/\Delta\lambda$ were obtained from
linear fits to the phase slope measurements in three bands by
\cite{2007AJ....133...26R}.  The scatter in Fig.~\ref{Fig.DBetaDLambdaVsAlbedo}
is substantial and no clear relation is evident between albedo and the
wavelength dependance of the phase function.  As a result, the interpretation
of these results in terms of physical properties of the surface is problematic.
The photometric (and polarimetric) phase functions depend in a non-trivial way
on the size and spatial arrangement of surface regolith particles, as well as
on their composition.  Besides, the phase functions of KBOs can only be
measured in a narrow range of phase angles (currently $0.5<\alpha<1.2$ deg in
the case of Haumea) making it difficult to recognize the presence or width of
narrow opposition peaks.  Nevertheless, further evidence that coherent
backscattering is responsible for the observed linearity between $\beta$ and
$\lambda$ can be sought using polarization measurements of the surface of
Haumea.

\subsection{Satellite}

We stacked our $\sim$70 frames in each filter to increase the signal-to-noise
ratio of any real features around Haumea and so attempt to detect the
satellites.  In the stacked frames, we used a field star as representative of
the point-spread function (PSF), scaled it to the brightest pixel of Haumea and
subtracted it from the KBO.  The result is shown in Figs.~\ref{Fig.SatelliteJ}
and \ref{Fig.SatelliteH}, where the original image, the PSF, and the subtracted
image are plotted;  one satellite, Hi'iaka, is clearly visible through both
filters.  We measure a Hi'iaka-Haumea separation $d=0.872\pm0.002$\arcsec\
($\mathrm{PA}=184.67\pm0.13$ deg), and flux ratios (with respect to Haumea) of
$0.088\pm0.001$ and $0.064\pm0.002$ in $J$ and $H$, respectively.  The mean UT
of the frame stack is 2008 Apr 15.23134.  We do not detect any other sources in
the vicinity of Haumea, although we are sensitive to objects as faint as
0.11$\times$ the $J$-band (and 0.21$\times$ the $H$-band) flux of Hi'iaka.  At
the time of observation, the fainter, inner satellite Namaka was within
0.2\arcsec\ of Haumea (Ragozzine, private comm.), which explains why it is
invisible in our data.  Given its fractional brightness ($\sim$1\%) with
respect to Haumea, Namaka has negligible contribution to the near-infrared
photometry presented here.  

\FigSatelliteJ

\FigSatelliteH

Haumea has a color index $J\!-\!H=-0.057\pm0.016$ mag (see
Fig.~\ref{Fig.ColorCurves} and Table~\ref{Table.Colors}).  Our photometry of
Hi'iaka relative to Haumea implies a color $J\!-\!H=-0.399\pm0.034$ mag for the
former.  This color index is unusually blue, but consistent with the
observation that the 1.5 \mum\ band is deeper for Hi'iaka than for Haumea
\citep{2006Takato,2006ApJ...640L..87Bar}.  Table~\ref{Table.Colors} lists the
derived colors of Haumea and Hi'iaka.  A sample of $J\!-\!H$ colors of 40 KBOs
and Centaurs found in the literature \citep{2006AJ....131.1851Del} shows a
pronounced clustering around $J\!-\!H=0.4$ mag with a dispersion of $\sim$0.2
mag.  The two significant outliers are KBOs (19308) 1996 TO$_{66}$
($J\!-\!H=-0.21\pm0.17$ mag) and (24835) 1995 SM$_{55}$ ($J\!-\!H=-0.49\pm0.06$
mag) which both possess near-infrared spectra consistent with water-ice
absorption \citep{1999ApJ...519L.101B,2008AJ....135...55B}.

The $J\!-\!H$ colors of Haumea and Hi'iaka are significantly ($>$9$\sigma$)
different.  Given the current best estimates for the mass and orbit of Hi'iaka,
particles collisionally ejected (ejection velocities 10 to 200 m~s$^{-1}$) from
its surface will likely reach Haumea only on hyperbolic orbits
\citep{2008arXiv0805.3482Ste}.  The escape speed from the surface of Hi'iaka
(near 130 m~s$^{-1}$, assuming water-ice density) rivals the escape speed from
the Haumea system at the orbit of Hi'iaka ($\sim$120 m s$^{-1}$), meaning that
only hyperbolic mass exchange is possible.  The same cannot be said about
Namaka which is both smaller and deeper into the potential well of the
satellite system.  It is therefore more likely that Haumea is polluted by
ejecta from Namaka than from Hi'iaka.  Whether this means that the $J\!-\!H$
color of Namaka is closer to that of Haumea remains to be seen.

\section{Summary}

From time-resolved, near-infared photometry of Kuiper belt object Haumea we
find the following

\begin{itemize}

\item The near-infrared peak-to-peak photometric range is $\Delta m_R$ =
0.30$\pm$0.02 mag.  The new data reveal slight visible/near-infrared color
variations on Haumea, which are spatially correlated with a previously
identified surface region, redder in $B\!-\!R$ and darker than the mean
surface.  We find that near this region Haumea displays an enhanced $H$-band
reflectance accompanied by $B$-band absorption relative to elsewhere on the
surface.  Time-resolved spectra are needed to learn more about the
physicochemical properties of this anomalous region.

\item The rotationally medianed visible and near-infrared colors of Haumea are
$R\!-\!J=0.885\pm0.012$ mag and $J\!-\!H=-0.057\pm0.016$ mag.  

\item We detect Hi'iaka, Haumea's brightest satellite, in both $J$ and $H$ and
measure its color $J\!-\!H=-0.399\pm0.034$ mag.  The $J\!-\!H$ color difference
between Hi'iaka and Haumea is significant ($>$9$\sigma$).  This suggests that
either the transfer of surface ejecta between the two is negligible, or that
their surface colors are not controlled by ejecta transfer.  Ejecta transfer
between Haumea and the inner satellite Namaka is neither ruled out nor
substantiated by our data but is more likely given the configuration of the
system.

\item The slope of the $J$-band phase function in the range
$0.55\leq\alpha(\mathrm{deg})\leq1.14$ is $\beta_J=0.16\pm0.04$ mag deg$^{-1}$.
Combining this measurement with slopes obtained in three other visible
wavelengths we find that the slope of Haumea's phase function varies
monotonically with wavelength.  The slope of the relation is $\Delta
\beta/\Delta \lambda\sim 0.09$ mag deg$^{-1}$ $\mu$m$^{-1}$ and
$\beta_{1\mu\mathrm{m}}\sim 0.14$ mag deg$^{-1}$.  This finding confirms
previous inferences that coherent backscattering is the main cause of
opposition brightening for Haumea.\looseness-1

\end{itemize}

\section*{Acknowledgments}

We appreciate insightful discussion and comments from David Jewitt and Jan
Kleyna.  We thank Will Grundy for sharing useful software routines, and Jon
Swift for valuable discussion that contributed to this work.  PL was supported
by a grant to David Jewitt from the National Science Foundation.

\end{document}